# PREDICTING THE ACTIVITY OF CHEMICAL COMPOUNDS BASED ON MACHINE LEARNING APPROACHES

**Do Hoang Tu[1], Tran Van Lang[2*], Pham Cong Xuyen[1], Le Mau Long[1,3,]**

[1] Lac Hong University, Vietnam
[2] HCMC University of Foreign Languages - Information Technology, Vietnam
[3] Nguyen Tat Thanh University, Vietnam

*dhtus.vn@gmail.com, langtv@huflit.edu.vn, pcxuyen@lhu.edu.vn, lmaulong@gmail.com*

**ABSTRACT** — *Exploring methods and techniques of machine learning (ML) to address specific challenges in various fields is essential. In this work, we tackle a problem in the domain of Cheminformatics; that is, providing a suitable solution to aid in predicting the activity of a chemical compound to the best extent possible. To address the problem at hand, this study conducts experiments on 100 different combinations of existing techniques. These solutions are then selected based on a set of criteria that includes the G-means, F1-score, and AUC metrics. The results have been tested on a dataset of about 10,000 chemical compounds from PubChem that have been classified according to their activity.*

## I. INTRODUCTION

In datasets used in biological experiments for measuring the activity of various compounds against different biological targets, often used in screening, there is usually a significant imbalance between active and inactive compounds, with the number of inactive data points being much larger. Therefore, training requires the use of suitable machine learning models. Additionally, preprocessing before using machine learning methods for training is also a crucial issue. The following issues are approached to address the problem of predicting the activity of chemical compounds using chemistry-related datasets:

- Investigating the dependency of attributes or features in the dataset to potentially reduce the number of features. This can be done using methods such as ANOVA F-test to assess the dependency of each feature on the target variable or by using correlation coefficients.
- Data normalization to mitigate the variance of observations (dataset) within the same feature.
- Handling data imbalance using resampling techniques such as oversampling or undersampling.
- Combining grid search, cross-validation, and Bayesian optimization techniques to search for hyperparameters, including sensitive parameters like learning rate, epochs, batch size, and insensitive parameters to find suitable hyperparameters for the model.

## II. RELATED WORKS

Predicting the activity of chemical compounds is a crucial area in pharmacology and chemoinformatics research. Several noteworthy research works in this field are as follows. The paper "*A Deep Learning Approach to Antibiotic Discovery*" **[1]** explores the use of deep learning to predict the antibiotic activity of compounds, aiding in identifying compounds with antibacterial properties. The study experimented with over 107 million molecules from the ZINC15 database. The results identified eight antibacterial compounds with distinct structures from known antibiotics. This research highlights the role of deep learning methods in expanding the arsenal of antibiotics by discovering structurally different antibacterial molecules. It represents a significant breakthrough in antibiotic discovery, potentially aiding in finding new compounds to combat antibiotic-resistant bacteria. The paper "*Predicting Antitumor Activity of Peptides by Consensus of Regression Models Trained on a Small Data Sample*" **[2]** focuses on solving regression problems with a small dataset of only 429 compounds. The study employs various methods such as linear regression, polynomial regression, Gaussian kernel regression, neural networks, k-nearest neighbors (kNN), and support vector machines (SVM) to address overfitting. It develops a method to predict the anti-tumor activity of peptides using a consensus of regression models trained on a small dataset. This approach demonstrates higher accuracy compared to previous methods, even with limited data. It can be used to develop new anti-cancer drugs and streamline the screening of active peptides, saving research time and costs. To handle overfitting issues in the training process and improve model performance, the work "*Large-scale evaluation of k-fold cross-validation ensembles for uncertainty estimation*" **[3]** was published in the Journal of Cheminformatics in 2023. The paper "*Chemprop: A Machine Learning Package for Chemical Property Prediction*" **[4]** introduces a new software package called Chemprop. This software can be used to build and train machine learning models to predict chemical properties of compounds. Chemprop employs a novel artificial neural network architecture called Directed Message-Passing Neural Network (D-MPNN) and has been proven effective in predicting chemical properties of compounds. The paper evaluated Chemprop's effectiveness on a dataset comprising over 100,000 compounds, showing its ability to predict chemical properties with higher accuracy compared to other methods.

---

* Corresponding Author

In addition to research papers published in journals and conference proceedings, there are also software tools and applications harnessing machine learning algorithms to address the posed challenges. For instance, AlphaFold (https://alphafold.com) is software designed for predicting the 3D structure of proteins with remarkable accuracy. Understanding protein structure can aid in predicting protein-ligand interactions, guiding the design of active molecules. The DeepChem library (https://deepchem.io/about/) is an open-source library that utilizes deep learning to predict the properties and activities of molecules. This library itself supports various tasks, including predicting binding affinities or the potential activation of molecules.

Another noteworthy example is ChemBERTa (https://arxiv.org/abs/2209.01712), a pre-trained language model developed for predicting the properties and activities of molecules. These models can learn chemical structures from data and make predictions about essential properties. The results of this work were published in the paper "*ChemBERTa-2: Towards Chemical Foundation Models*" which was posted on arXiv on September 9, 2022. This research focuses on leveraging Large Language Models (LLMs) to learn chemical features from text. As a result, the authors created a new LLM called ChemBERTa-2, which was trained on a dataset consisting of over 77 million SMILES strings. SMILES is a string format used to represent chemical molecules **[5]**. The authors demonstrated that this approach can learn chemical features from text and can be used to perform various chemistry tasks, such as predicting biological activity, molecular structure, and chemical language translation. This work represents a significant step forward in using LLMs for chemical tasks. LLMs have the potential to automate many time-consuming and costly chemistry tasks and assist scientists in developing new drugs and materials.

In the field of Chemoinformatics, researchers often predict activity based on molecular structure using Quantitative Structure-Activity Relationship (QSAR) models. This approach involves predicting activity based on the relationship between molecular structure and activity. In essence, it employs statistical models to forecast the activity of chemical compounds based on their structure.

However, the prediction of the activity of chemical compounds continues to evolve with the integration of various methods and technologies. The research works mentioned above represent only a small part of the series of studies that have been and are currently being conducted in this field. In the paper [5] on the advances and challenges of machine learning in Chemoinformatics, the authors also addressed some challenging problems that arise in Chemoinformatics when leveraging machine learning techniques and methods. These are issues that require the involvement of the community of data scientists and computer scientists to enhance the accuracy of computational approaches to address the challenges posed by Chemistry.

Research efforts in predicting the activity of chemical compounds have made significant progress in recent years. The results obtained from these efforts have been used to develop new drugs, new materials, and new catalysts. However, the methods and techniques in Computer Science and Data Science still have limitations to overcome in achieving comprehensive solutions in this field.

With the issue at hand and the related works presented in the two sections above related to predicting the activity of chemical compounds. To be more specific, Part III is used to present the approach methods. The experimental results are presented in Part IV, and the final section is a concluding commentary on the use of machine learning in addressing the classification of activity in chemical compounds.

## III. APPROACH METHOD

With the development of technology, today's methods, machine learning techniques, and supporting tools have become quite diverse. Understanding how to use them is a quick approach to solving the posed problem.

### A. Investigating the Dependency among Features

Trong công trình này, hệ số tương quan (*correlation coefficient*) được sử dụng để rút gọn thuộc tính thông qua việc chọn lọc đặc trưng. Hệ số tương quan giữa 2 thuộc tính $x$ và $y$ có $N$ mẫu (*data point*) tương ứng là $(x_i, y_i)$, $\forall i = \overline{1, N}$ được tính qua công thức:

In this work, the correlation coefficient is used to reduce attributes through feature selection. The correlation coefficient between two features, $x$ and $y$, in a dataset with $N$ corresponding data points $(x_i, y_i)$, $\forall i = \overline{1, N}$, is calculated using the formula:

$$r_{xy} = \frac{\sum_{i=1}^{N}(x_i - \overline{x})(y_i - \overline{y})}{\sqrt{\sum_{i=1}^{N}(x_i - \overline{x})^2}\sqrt{\sum_{i=1}^{N}(y_i - \overline{y})^2}}$$

Where $\overline{x} = \frac{1}{N}\sum_{i=1}^{N} x_i$ và $\overline{y} = \frac{1}{N}\sum_{i=1}^{N} y_i$.

When the correlation coefficient is +1 or -1, it indicates a perfect positive or negative correlation, respectively.

To perform feature selection using the correlation coefficient, you can follow these steps using existing libraries:

- Calculate the correlation matrix by using the corr() function in Pandas. This will compute the correlation matrix of the attributes within a DataFrame.

```
corr_matrix = df_train.drop('Target', axis=1).corr()
corr_matrix = df_train.drop('Target', axis=1).corr()
```

- Define a threshold: Determine a threshold value to decide whether to keep or discard attributes. The choice of threshold depends on the specific problem and the importance of attributes. Typically, a threshold of 0.75 is used when considering the absolute values of correlation coefficients.
- Attribute selection: Based on the correlation matrix and the chosen threshold, select the attributes that have correlation coefficients below the threshold. This means that you retain only the necessary attributes while discarding those that are highly correlated with others.

```
threshold = 0.75
corr_features = set()

for i in range(len(corr_matrix.columns)):
    for j in range(i):
        if abs(corr_matrix.iloc[i, j]) > threshold:
            colname = corr_matrix.columns[i]
            corr_features.add(colname)

selected_features = set(df_train.columns) - corr_features
```

Reducing attributes based on correlation can indeed be a simple method, but it requires careful consideration and experimentation to ensure that the retained attributes still contain important information in the data. Sometimes, a feature may have low correlation with the class label but could carry valuable information for classification. Additionally, the correlation between features may not always be easily measured using Pearson correlation coefficient alone. Sometimes, having correlations between certain features in a multidimensional space can be more meaningful. Reducing attributes based on correlation can introduce multicollinearity issues, where some features are removed, and the correlation between the remaining features changes. This can impact the model's interpretability and understanding. It's indeed a challenge when dealing with data preprocessing in the field of Cheminformatics. To address these challenges, it's important to:

- Carefully Define Objectives: Clearly define your objectives and what you aim to achieve by reducing attributes. Make sure you understand the trade-offs between reducing dimensionality and preserving important information.
- Experiment and Validate: Experiment with different correlation thresholds and evaluate the impact on model performance. Use cross-validation to validate the effectiveness of feature selection.
- Consider Domain Knowledge: Incorporate domain knowledge if available. Sometimes, domain expertise can help in identifying which features are truly important even if they have lower correlations.
- Use Advanced Techniques: Consider more advanced dimensionality reduction techniques, such as Principal Component Analysis (PCA) or t-Distributed Stochastic Neighbor Embedding (t-SNE), which can capture complex relationships between features.
- Monitor Model Performance: Continuously monitor the performance of your machine learning model to ensure that it's not adversely affected by attribute reduction.
- Visualize the Data: Visualize the data and the relationships between features using techniques like scatter plots or dimensionality reduction visualizations to gain insights.
- Handle Multicollinearity: If multicollinearity becomes a concern, you can explore techniques like ridge regression or Lasso regression, which can handle multicollinearity in the modeling process.

These are the unresolved issues in this work.

### B. Handling Data Imbalance

### 1. Sampling Techniques

To enhance the dataset, one can employ methods such as SMOTE and ADASYN. Specifically, SMOTE (Synthetic Minority Over-sampling Technique) is a commonly used method to augment data. It generates additional samples for the minority class by sampling neighboring minority samples and creating a new sample by averaging them.

On the other hand, ADASYN (Adaptive Synthetic over-sampling method) is also a technique to create new samples for the minority class based on the level of data imbalance. It focuses on generating new samples in low-density areas of the minority class. This can help improve the performance of machine learning models on severely imbalanced datasets.

SMOTE and ADASYN are two of the most popular techniques for addressing data imbalance. They both create new samples by resampling minority samples and generating synthetic samples from them. However, there are some key differences between these two techniques. SMOTE creates new samples by sampling neighboring minority samples and creating a new sample by averaging the selected samples. This can sometimes lead to the generation of unrealistic samples. ADASYN, on the other hand, creates new samples by focusing on generating samples in low-density regions of the minority class. This can help improve the performance of machine learning models on datasets with severe class imbalance.

Additionally, there are other techniques like Borderline-SMOTE, SMOTETomek, and SMOTE-ENN, which are newer methods developed based on SMOTE. They all aim to improve SMOTE's performance by addressing some of its limitations. For example, Borderline-SMOTE focuses on creating new samples in the boundary regions of the minority class. This can help improve the performance of machine learning models on datasets with severe class imbalance.

SMOTETomek combines both SMOTE and Tomek Links. Tomek Links is a method for removing close points between different classes to create a clearer boundary between classes. When combined with SMOTE, SMOTETomek generates new samples in the space between classes and then removes close points between classes. However, if the original data does not contain strong noise or close points between classes, the results of SMOTE and SMOTETomek may become similar. This can happen when oversampling and removing close points do not significantly change the class imbalance situation.

SMOTE-ENN combines SMOTE and Edited Nearest Neighbors (ENN). ENN is a technique for removing samples whose nearest neighbors belong to the opposite class. This can help improve the performance of machine learning models on imbalanced datasets by removing samples that can confuse the model.

SVMSMOTE is a data augmentation method for imbalanced data based on the SVM algorithm. It generates additional samples for the minority class based on the near-boundary space of that class. SVMSMOTE operates by identifying the near-boundary space of the minority class using an SVM trained on the original data to determine support vectors and the near-boundary space of the minority class. Then, minority samples near the near-boundary space are selected to create new samples. This can help improve the performance of machine learning models by making the minority class more important and achieving a better balance in class sizes.

To address severe data imbalance in a dataset, Python libraries such as the imbalanced-learn library can be used. This is a powerful Python library for handling severe class imbalance. It offers various techniques to address class imbalance, including oversampling, undersampling, and ensemble learning methods. The library focuses on modifying the training set to achieve a balanced distribution between classes.

For using SMOTE to augment training data:

```
smote = SMOTE(random_state=42)
X_train_smote, y_train_smote = smote.fit_resample(X_train, y_train)
print("Số lượng mẫu sau khi resample:", len(X_train_smote))
```

Utilize ADASYN to further enhance the dataset:

```
adasyn = ADASYN()
X_train_adasyn, y_train_adasyn = smote.fit_resample(X_train, y_train)
print("Số lượng mẫu sau khi resample:", len(X_train_adasyn))
```

## 2. Using the GAN Model

A GAN (Generative Adversarial Network), also known as a generative adversarial model or simply a GAN model, is a machine learning model that uses two competing machine learning models. The first model, called the 'generator,' aims to generate new samples from input data. The second model, called the 'discriminator,' classifies samples as real or fake. These two models are trained together, where the 'generator' tries to create samples that the 'discriminator' cannot distinguish from real data, and the 'discriminator' tries to accurately classify samples as real or fake.

The GANs have proven to be very effective in generating new data samples that resemble real data. This has made them popular in various applications, including handling data imbalance. In data imbalance handling, GANs can be used to generate new samples for minority classes. This can help improve the performance of machine learning models on imbalanced datasets by making the classes more balanced. GANs have been shown to be effective in improving the performance of machine learning models on imbalanced datasets. However, GANs also have a fundamental limitation, which is that they can generate unrealistic new samples.

The implementation steps are as follows:

- Build a GAN model:

Create the Generator

```
generator = Sequential()
generator.add(Dense(32, input_dim=dim, activation='relu'))
generator.add(Dense(dim, activation='sigmoid'))
generator.compile(loss='binary_crossentropy', optimizer=Adam())
```

And the discriminator, with `kernel_initializer=initializers.RandomNormal(stddev=0.02)` to determine the initialization method for weight matrices of the layers in the discriminator network using a standard distribution (Random Normal Disttibution) with a standard deviation of 0.02.

```
discriminator = Sequential()
```

```
discriminator.add( Dense(32, input_dim=dim, activation='relu',
            kernel_initializer = initializers.RandomNormal(stddev=0.02)) )
discriminator.add( Dense(1, activation ='sigmoid',
            kernel_initializer = initializers.RandomNormal(stddev=0.02)))
discriminator.compile(loss='binary_crossentropy', optimizer=Adam(), metrics=['accuracy'])
```

Create a GAN model based on the Generator and Discriminator:

```
gan = Sequential()
gan.add(generator)
gan.add(discriminator)
gan.compile(loss='binary_crossentropy', optimizer=Adam())
```

- Train the GAN model by generating fake data, combining real and fake data, training the discriminator, and finally, training the generator.

```
for epoch in range(epochs):
        random_noise = np.random.normal(0, 1, size=[count_minority,dim])
        generated_data = generator.predict(random_noise)

        X_combined = np.concatenate([X_train, generated_data])
        y_combined = np.concatenate([y_train, np.zeros(count_minority)])

        discriminator.trainable = True
        discriminator.fit(X_combined,y_combined,batch_size=batch_size,epochs=1,verbose=0)

        discriminator.trainable = False
        gan.fit(random_noise,np.ones(count_minority),batch_size=batch_size,epochs=1, verbose=0)
```

- Generate new dataset from the trained generator

```
random_noise = np.random.normal(0,1,size=[count_minority, X_train.shape[1]] )
generated_data = generator.predict(random_noise)
```

- Combine the new data with the original data:

```
X_augmented = np.concatenate([X_train, generated_data])
y_augmented = np.concatenate([y_train, np.zeros(count_minority)])
```

## 3. Ensemble Learning

Ensemble learning is a machine learning technique that combines the results of multiple different or similar base models to create a stronger model. Base models in ensemble learning can be either different machine learning models or identical models trained on different datasets. This leverages the diversity and knowledge aggregation from various models to improve the predictive capabilities of the final model. Ensemble learning can also be used to enhance the performance of machine learning models on imbalanced datasets. Ensemble learning techniques are typically categorized into two main types:

- Bagging (Bootstrap Aggregating): This method involves creating multiple sub-datasets by sampling with replacement from the original training data. These sub-datasets are randomly sampled with replacement, meaning that a sample can appear multiple times in a sub-dataset. Each sub-dataset is then used to train a separate predictive model. The predictions from these individual models are then combined, for example, by taking the average (for regression problems) or by voting (for classification problems). The Bagging technique can help improve model performance by reducing bias and enhancing the stability of models. Random Forest is a well-known example of a Bagging technique.
- Boosting: This method focuses on creating sequentially predictive models, where each subsequent model focuses on correcting errors made by the previous one. Predictive models are generated based on data weights, where minority class samples are assigned higher weights, while majority class samples are assigned lower weights. Boosting can help improve model performance by focusing on minority class samples and improving the discrimination between minority and majority class samples. Data points misclassified by the model are assigned higher weights for the next model to address. AdaBoost and Gradient Boosting are common examples of Boosting techniques.

In the case of imbalanced data, when one class has more samples than the other, several useful ensemble learning techniques have been implemented in the imbalanced-learn library, including:

- EasyEnsembleClassifier: As proposed in **[6]**, this algorithm aims to overcome the imbalanced data drawback. Easy Ensemble samples multiple small subsets from the majority class, trains a learner using each subset, and combines the results of those learners. Experimental results have shown that this technique performs well compared to others at that time. It creates multiple subsets from the minority class to address class imbalance by reducing the number of samples from the majority class. Specifically, EasyEnsembleClassifier divides the data into small subsets with the same number of samples as the minority class. Each subset is then used to train an individual classification model, such as a Decision Tree or Random Forest. Finally, the individual models are combined to form an ensemble model. The goal of EasyEnsemble is to create classification models capable of handling imbalanced data and improving classification performance on the minority class. By creating multiple models from data subsets, EasyEnsembleClassifier helps reduce overfitting on minority data and provides accurate predictions for both classes.

- BalancedRandomForestClassifier: This class combines Bagging and the use of different data weights to create a more powerful classification model. BalancedRandomForestClassifier works by creating multiple sub-datasets through random sampling with replacement from the original training data, with data augmentation performed by assigning weights to the data points during the creation of each sub-dataset. These weights focus on balancing the data classes, giving higher importance to minority class data. This helps the model concentrate more on accurately classifying minority data points. Finally, it integrates individual decision trees trained on these sub-datasets by taking a majority vote to create the final classification model.

- BalancedBaggingClassifier: In addition to BalancedRandomForestClassifier, there is also the BalancedBaggingClassifier class, which offers more flexibility regarding the choice of base models, allowing customization based on specific problem requirements. While BalancedRandomForestClassifier is limited to using decision trees as the base model, suitable for scenarios where decision trees perform well, BalancedBaggingClassifier is a more generalized approach. It combines the strengths of BalancedRandomForestClassifier while leveraging the advantages of EasyEnsembleClassifier for a broader range of applications.

## 4. Using Loss Function

Loss functions can also be used to address the issue of class imbalance in classification problems. A loss function is an evaluation function that assesses the performance of a machine learning model. In the case of class imbalance, loss functions can be employed to make the model focus more on minority samples. Some common loss functions used to handle imbalanced data include:

- Balanced Cross-Entropy Loss: This loss function balances the weights of errors for minority and majority samples. The function is generally formulated as follows:

$$L_{bce} = -\frac{1}{N}\sum_{i=1}^{N} y_i \log \widehat{y_i} + (1 - y_i) \log(1 - \widehat{y_i})$$

Where $N$ is the number of samples in the dataset, $y_i$ is the actual label of the $i$-th sample, and $\widehat{y_i}$ is the model's prediction for this sample $i$-th.

- Weighted Cross-Entropy Loss: This function assigns different weights to errors for minority and majority samples. The function is formulated as follows:

$$L_w = -\frac{1}{N}\sum_{i=1}^{N} w_i y_i \log \widehat{y_i} + (1 - y_i) \log(1 - \widehat{y_i})$$

Where $w_i$ is weight of the $i$-th sample.

- Focal Loss: This function focuses on hard-to-classify samples, and it takes the form of:

$$L_f = -\frac{1}{N}\sum_{i=1}^{N} y_i \log \widehat{y_i} + (1 - y_i) \log(1 - \widehat{y_i}) \cdot \gamma^{1-y_i}$$

Where, $\gamma_i$ is a tuning parameter.

- Dice Loss: This loss function is calculated by taking the product of the Dice similarity and the Jaccard similarity. The Dice similarity is computed by taking the sum of the intersections of two sets of samples and dividing it by the sum of the areas of the two sample sets. The Jaccard similarity is calculated by taking the intersection of two sample sets and dividing it by the total area of the two sample sets. The function with Dice similarity has the following form:

$$L_d = 1 - \frac{2\sum_{i=1}^{N} y_i \widehat{y_i}}{\sum_{i=1}^{N} y_i + \sum_{i=1}^{N} \widehat{y_i}}$$

The function with Jaccard similarity has the following form:

$$L_j = 1 - \frac{\sum_{i=1}^{N} y_i \widehat{y_i}}{\sum_{i=1}^{N} y_i + \sum_{i=1}^{N} \widehat{y_i} - \sum_{i=1}^{N} y_i \widehat{y_i}}$$

To implement this, we can use the Keras library with a deep neural network model to utilize these custom loss functions.

# IV. EXPERIMENTAL RESULTS

## A. Dataset

To conduct the experiments, we used the bioassay screening dataset available on Kaggle at the following address: https://www.kaggle.com/datasets/uciml/bioassay-datasets. This dataset is from PubChem and has also been uploaded to the UC Irvine machine learning data repository [7]. The dataset is a collection of 21 biological assays (screens) measuring the activity of various compounds against different biological targets. In this article, we experimented with the AID373red_train.csv sample. AID373 is a simulated dataset for classifying chemical compounds based on their activity against intracellular targets. It contains 47,831 compound samples, with 50 active samples and 47,781 inactive samples; a significant class imbalance when used as a dataset for training with two labels, active and inactive. This dataset consists of 153 columns of data, including IDs for identification purposes. These columns represent various chemical properties of the compounds. The dataset was collected by the Molecular Screening Center at the Scripps Research Institute to develop machine learning models that can predict the activity of chemical compounds for various purposes, such as disease treatment. It is the largest and most diverse dataset of its kind. The columns have the following values:

- The first columns (from column 1 to column 121): The values in these columns are all 0 and 1, representing binary features.
- The next columns (from column 122 to column 146): The values in these columns are real numbers, containing information about chemical and physical features of the compounds.
- Columns from 147 to 153 are numerical values, representing PSA, NumRot, NumHBA, NumHBD, MW, BBB, and BadGroup. PSA is the polar surface area of the compound, which measures the size and polar nature of the compound's outer surface. Compounds with a higher PSA may dissolve better in water and can easily penetrate cells. NumRot is the number of rotatable bonds in the compound. Rotatable bonds are bonds that can freely rotate around their axis. Compounds with more rotatable bonds tend to be less stable and more prone to degradation. NumHBA is the number of hydrogen bond acceptors in the compound. Hydrogen bond acceptors are molecules with hydrogen atoms that can bond with atoms of higher electronegativity, such as oxygen and nitrogen. Hydrogen bond acceptors can form hydrogen bonds with hydrogen bond donors, which may help them dissolve in water and penetrate cells. NumHBD is the number of hydrogen bond donors in the compound. Hydrogen bond donors are molecules with hydrogen atoms that can bond with atoms of higher electronegativity, such as oxygen and nitrogen, forming hydrogen bonds. MW is the molecular weight of the compound, which is a measure of the compound's mass. Compounds with a higher molecular weight may be less soluble in water and less able to penetrate cells. BBB is the ability of the compound to cross the blood-brain barrier. The blood-brain barrier is a barrier that separates the blood from the cerebrospinal fluid. Compounds with better blood-brain barrier penetration ability can enter the brain and treat neurological diseases. BadGroup represents the hazardous group of the compound. Hazardous groups are a group of molecules that can be harmful to the human body. Compounds with hazardous groups may not be safe for use as drugs.
- The last columns (column 154) are character strings representing the activity status of the sample, with "Active" and "Inactive" indicating whether the sample has biological activity or not. This column serves as the target label indicating the activity status of the sample.

## B. Scoring Metrics

The G-mean (Geometric Mean) is a statistical metric used to evaluate the performance of classification models, especially when dealing with imbalanced data. The G-mean is calculated as:

$$G - mean = \sqrt{TPR \times TNR}$$

Where $TPR$ (True Positive Rate) is also known as recall, representing the percentage of true positive samples that are correctly predicted $TPR = \frac{TP}{TP+FN}$. And TNR (True Negative Rate) is also known as specificity, representing the percentage of true negative samples that are correctly predicted, $TNR = \frac{TN}{TN+FP}$.

The G-mean measures a model's ability to correctly predict both positive and negative classes while considering the balance between the two classes. Additionally, the F1-score is another metric that combines sensitivity and specificity, both of which are important metrics. Sensitivity indicates how accurately the model can classify minority class data points, while specificity indicates how accurately the model can classify majority class data points. The F1-score balances sensitivity and specificity and is a good metric for evaluating model performance on imbalanced data. The F1-score is calculated as:

$$F1 - score = 2\frac{TPR \times TNR}{TPR + TNR}$$

Both the G-mean and F1-score are good performance metrics to use when dealing with imbalanced data. However, they have different strengths and weaknesses.

The G-mean is an imbalanced performance metric, meaning it is not influenced by the number of samples in each class. This makes it a good choice for classification tasks that focus on detecting minority class samples.

The F1-score is a balanced performance metric, meaning it balances sensitivity and specificity. This makes it a good choice for classification tasks aiming to achieve a balance between detecting minority class samples and not misclassifying majority class samples.

In addition to these metrics, the AUC (Area Under the Curve) is another imbalanced performance metric. It is commonly used for binary classification problems where data can be imbalanced. The formula for calculating AUC involves finding the area under the curve of the TPR plotted against the FPR (False Positive Rate):

$$AUC = \int_0^1 TPR(FPR)dFPR$$

Where FPR (False Positive Rate) is the ratio of false positives to the total number of actual negatives: $FPR = \frac{FP}{TN+FP}$.

An AUC value of $\geq 0.7$ suggests that the model can be used for classification; however, AUC is not a perfect performance metric.

Therefore, combining these prioritized metrics into an evaluation score is necessary. In this work, the highest priority is given to G-mean, followed by F1-score, and finally, AUC, as described in the formula provided.:

$$Score = G - mean \times 3 + F1 - score \times 2 + AUC \times 1$$

### C. Results

#### 1. Preprocessing

To reduce the number of attributes based on correlation coefficients, as previously mentioned. The results involve the removal of 19 attributes with correlation coefficients above a commonly used threshold of 0.75. Experimental results across various models showed inconsistent outcomes on the Chemistry dataset. Some models performed very well after attribute reduction, while others did not. This indicates that attribute reduction based solely on examining correlation coefficients with a chosen threshold may not be an optimal solution if it only considers attribute statistics without considering their inherent characteristics.

Illustrating with Ada Boost as in Figure 1b, the results are not favorable after attribute reduction (with a Score of 0.908), whereas without attribute reduction, the Score is 3.345 (Figure 1a). However, with Balanced Bagging, the opposite is observed; after attribute reduction, the Score is 4.28, whereas without attribute reduction, it is only 3.07.

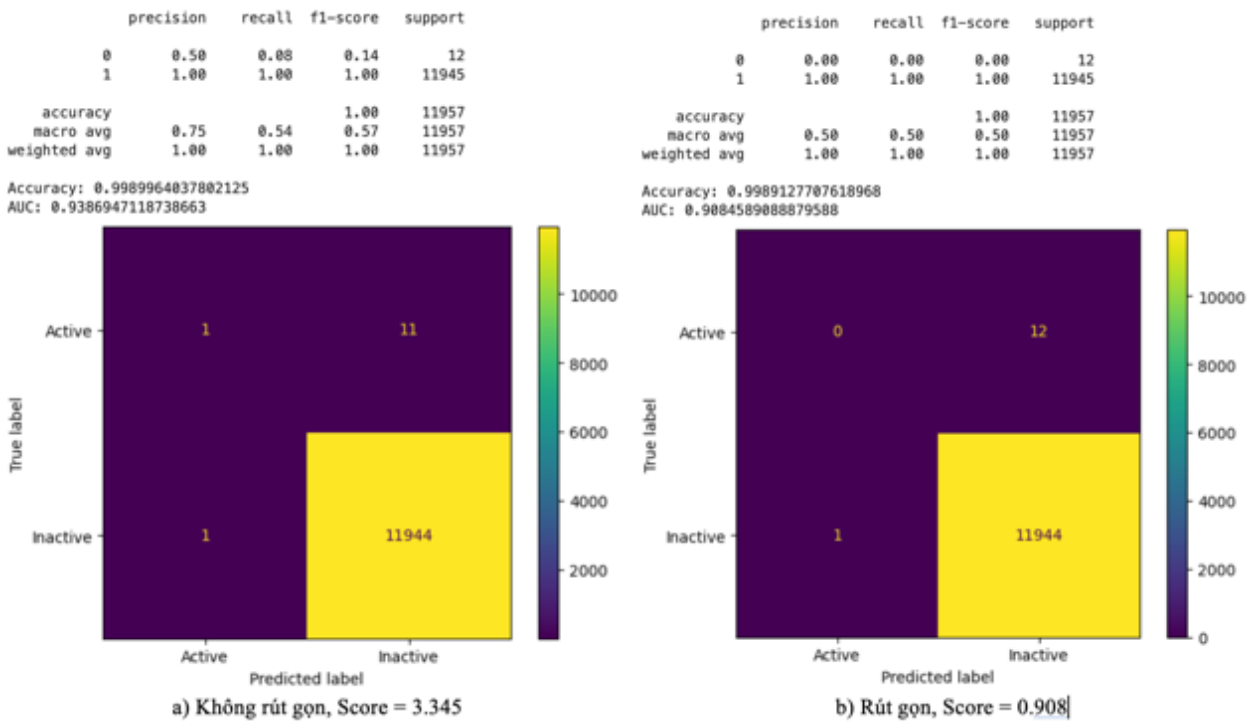

Figure 1. Comparing Attribute Reduction with Ada Boost

The data preprocessing step before model training also involves the application of Min-Max Scaling technique. All the experimental results are available in the Excel file at the following address: https://github.com/langtv/FAIR2023

## 2. Experimental through various Approaches

The experimental results encompassed 100 approaches by combining the techniques presented in the previous section. For parameter selection, an optimization technique with Bayesian Optimization (BO) was utilized. The list of the top 20 approaches by Score is provided in Table 1.

Table 1.   List of the Top 20 Approaches Ranked by High Score

| Approached Methods | G-mean | AUC | F1-score | Pretict: Active | | Predict: Inactive | | Score |
|---|---|---|---|---|---|---|---|---|
| | | | | True | False | True | False | |
| Balanced Bagging not majority (rút gọn thuộc tính) | 0,9996 | 0,71 | 0,29 | 2 | 10 | 11945 | 0 | 4,28 |
| AdaBoost_47831 | 0,7068 | 0,94 | 0,14 | 1 | 11 | 11944 | 1 | 3,34 |
| Balanced Random Forest not majority Borderline-SMOTE | 0,7068 | 0,66 | 0,25 | 2 | 10 | 11943 | 2 | 3,28 |
| Balanced Random Forest not majority | 0,7068 | 0,71 | 0,14 | 1 | 11 | 11944 | 1 | 3,11 |
| Balanced Bagging not majority | 0,6322 | 0,71 | 0,24 | 2 | 10 | 11942 | 3 | 3,07 |
| Balanced Random Forest not majority Borderline-SMOTE (rút gọn thuộc tính) | 0,5771 | 0,70 | 0,13 | 1 | 11 | 11943 | 2 | 2,70 |
| Balanced Bagging not majority (rút gọn thuộc tính) | 0,4998 | 0,66 | 0,13 | 1 | 11 | 11942 | 3 | 2,41 |
| Balanced Random Forest not majority SMOTE | 0,4262 | 0,78 | 0,17 | 2 | 10 | 11936 | 9 | 2,40 |
| Balanced Random Forest not majority SMOTE Tomek | 0,4262 | 0,78 | 0,17 | 2 | 10 | 11936 | 9 | 2,40 |
| Balanced Random Forest not majority SMOTE (rút gọn thuộc tính) | 0,4470 | 0,69 | 0,18 | 2 | 10 | 11937 | 8 | 2,40 |
| Balanced Random Forest not majority SMOTE Tomek (rút gọn thuộc tính) | 0,4470 | 0,69 | 0,18 | 2 | 10 | 11937 | 8 | 2,40 |
| Balanced Random Forest not majority ADASYN (rút gọn thuộc tính) | 0,4081 | 0,77 | 0,17 | 2 | 10 | 11935 | 10 | 2,33 |
| Balanced Bagging not majority Borderline-SMOTE (rút gọn thuộc tính) | 0,4262 | 0,58 | 0,17 | 2 | 10 | 11936 | 9 | 2,21 |
| Balanced Bagging not majority SVM SMOTE (rút gọn thuộc tính) | 0,4081 | 0,62 | 0,11 | 1 | 11 | 11940 | 5 | 2,07 |
| DeepLearning, BO, GAN, SMOTETomek_95562 | 0,3014 | 0,83 | 0,12 | 2 | 10 | 11925 | 20 | 1,97 |
| DeepLearning, BO, GAN, BorderlineSMOTE_95562 | 0,2948 | 0,83 | 0,11 | 2 | 10 | 11924 | 21 | 1,94 |
| Balanced Bagging not majority SVM SMOTE | 0,3534 | 0,62 | 0,10 | 1 | 11 | 11938 | 7 | 1,88 |
| Balanced Bagging not majority Borderline-SMOTE | 0,3243 | 0,58 | 0,13 | 2 | 10 | 11928 | 17 | 1,81 |
| DeepLearning, BO, RandomOverSamples_95562 | 0,2772 | 0,76 | 0,11 | 2 | 10 | 11921 | 24 | 1,80 |
| DeepLearning, BO, SMOTETomek_95562 | 0,2721 | 0,75 | 0,10 | 2 | 10 | 11920 | 25 | 1,77 |

To select models suitable for classification, one should rely on the AUC score as mentioned above while also eliminating models with a significantly low correct prediction rate for "Active." Acceptable approaches can be found in Table 2.

Table 2.   List of acceptable and usable solutions.

| Approached Methods | G-mean | AUC | F1-score | Predict: Active | | Precict: Inactive | | Score |
|---|---|---|---|---|---|---|---|---|
| | | | | True | False | True | False | |
| Balanced Bagging not majority (Feature reduction) | 0,9996 | 0,71 | 0,29 | 2 | 10 | 11945 | 0 | 4,28 |
| Balanced Bagging not majority | 0,6322 | 0,71 | 0,24 | 2 | 10 | 11942 | 3 | 3,07 |
| Balanced Random Forest not majority SMOTE | 0,4262 | 0,78 | 0,17 | 2 | 10 | 11936 | 9 | 2,40 |
| Balanced Random Forest not majority SMOTE Tomek | 0,4262 | 0,78 | 0,17 | 2 | 10 | 11936 | 9 | 2,40 |
| Balanced Random Forest not majority ADASYN (Feature reduction) | 0,4081 | 0,77 | 0,17 | 2 | 10 | 11935 | 10 | 2,33 |

| Approached Methods | G-mean | AUC | F1-score | Predict: Active | | Precict: Inactive | | Score |
|---|---|---|---|---|---|---|---|---|
| | | | | True | False | True | False | |
| DeepLearning, BO, GAN, SMOTETomek_95562 | 0,3014 | 0,83 | 0,12 | 2 | 10 | 11925 | 20 | 1,97 |
| DeepLearning, BO, GAN, BorderlineSMOTE_95562 | 0,2948 | 0,83 | 0,11 | 2 | 10 | 11924 | 21 | 1,94 |
| DeepLearning, BO, RandomOverSamples_95562 | 0,2772 | 0,76 | 0,11 | 2 | 10 | 11921 | 24 | 1,80 |
| DeepLearning, BO, SMOTETomek_95562 | 0,2721 | 0,75 | 0,10 | 2 | 10 | 11920 | 25 | 1,77 |

The results show that when using a combination technique with Balanced Bagging, applying data augmentation without enhancing the majority class, and utilizing a decision tree with the DecisionTreeClassifier() class as the base estimator yielded the most prominent outcome. The entire source code, along with the experimental results, is provided on GitHub at the following address: https://github.com/langtv/FAIR2023.

## V. CONSLUSION

The problem of finding a highly accurate activity prediction model from chemical datasets is a challenging task to achieve good results when the class imbalance is significant. The techniques implemented through Python libraries have not completely addressed this issue. Even the use of approaches like data generation through GAN models or handling it through loss functions by altering weights or adding regularization parameters are research challenges that need to be addressed. Additionally, leveraging technologies implemented through libraries is an effective solution in today's era when dealing with real-world problems using machine learning approaches. For the challenges posed by chemical datasets, understanding the nature of the attributes within the datasets is a particularly important issue..